\documentclass[prd,showpacs,showkeys,10pt]{revtex4}
\usepackage{amsmath,amsthm,amsfonts,amssymb}
\usepackage[mathcal]{eucal}
\usepackage{mathrsfs}
\usepackage{graphicx}
\usepackage{dcolumn}
\usepackage{latexsym}
\usepackage{epsfig}
\usepackage{bm}
\usepackage[all]{xy}
\newcommand{\ie}{\begin{equation}}
\newcommand{\fe}{\end{equation}}

\newcommand\fverb{\setbox\fverbbox=\hbox\bgroup\verb}
\newcommand\fverbdo{\egroup\medskip\noindent%
            \fbox{\unhbox\fverbbox}\ }
\newcommand\fverbit{\egroup\item[\fbox{\unhbox\fverbbox}]}
\newbox\fverbbox

\def\text#1{\mbox{#1}}

\begin{document}

\title{Scalar field localization on 3-branes placed at a warped resolved conifold }

\author{J. E. G. Silva and C. A. S. Almeida}
\affiliation{Departamento de F\'{i}sica - Universidade Federal do Cear\'{a} \\ C.P. 6030, 60455-760
Fortaleza-Cear\' {a}-Brazil}

\begin{abstract}
We have studied the localization of scalar field on a 3-brane embedded in a six dimensional warped
bulk of
the form $M_{4}\times C_{2}$, where $M_{4}$ is a 3-brane and $C_{2}$ is a 2-cycle of a six resolved conifold
$\mathcal{C}_{6}$ over a $T^{1,1}$ space. Since the resolved conifold is singularity-free in $r=0$ depending on a resolution parameter $a$, we have analyzed the
behavior of the localization of scalar field when we vary the resolution parameter. On one hand, this enable us
to study the effects that a singularity has on the field. On the other hand we can use the resolution parameter as
a fine-tuning between the bulk Planck mass and 3-brane Planck mass and so it opens a new perspective to extend the hierarchy problem. Using a linear and a nonlinear
warp factor, we have found that the massive and massless
modes are trapped to the brane even in the singular cone ($a\neq 0$).
We have also compared the results obtained in this geometry and those obtained in other six-dimensional models, as string-like geometry and cigar-like universe
geometry.
\end{abstract}


\maketitle

\section{Introduction}
\indent \indent Since the original Randall-Sundrum model \cite{Randall:1999vf,Randall:1999ee} many works intends to extend the localization
of various fields on a 3-brane embedded in the higher dimensional bulk. Besides the localization of gravity and
other fields, many models have been suggested for explain other physical problems, for instance, the small value of the cosmological
constant \cite{Chen:2000at}. In order to explain the geometry
used to localize the fields in noncompact extra dimensions, some authors have assumed that the 3-brane is
generated by a topological defect. In a six-dimensional bulk, Cohen and Kaplan \cite{Cohen:1999ia} have found such a
geometry generated
by a global string. In this context the geometry has cylindrical symmetry and a naked singularity at $r=0$, where the 3-brane is placed,
and also another singularity far from the origin.
Gregory \cite{Gregory:1999gv} has found a nonsingular string-like solution by adding a cosmological constant to the bulk and splitting the metric inside
and outside the core of the string defect. For the continuity boundary condition on the core, Gregory has found the phase space of solutions describing the stable and
unstable points. For a geometry generated by a local string-like defect, Gherghetta and Shaposhnikov \cite{Gherghetta:2000qi} have found a solution with negative
cosmological constant on bulk that has trapped gravity. Oda \cite{Oda:2000zc} has
extended this solution for a bulk builded from a warped product of a (p-1)-brane
and a $S^{n}$ sphere and he has studied the localization of many kind of fields. All solutions above assume that the transverse space has spherical
symmetry and the whole bulk has cylindrical symmetry. We have studied an extension of this approach for the localization of
scalar field where the transverse space has a conifold geometry whose singularity depends on a resolution
parameter.

The conifold here is a conical manifold $\mathcal{C}^{n}$ over a $X^{n-1}$ called a base space. $X^{n-1}$ is topological
equivalent to $S^{n-1}$ defined by the coset $X^{n-1}=SU(n-1)/SU(n)$ \cite{Candelas:1989js}. It has a naked singularity that
arises as an orbifold fixed point of the group $Z_{n}$, i.e., $\mathcal{C}^{n}=R^{n}/Z_{n}$. The conifold is an example of a Calabi-Yau space, a Ricci-flat
manifold that is a candidate to a internal space in compactification
of string theories. The conifold is a
generator of all \textit{Calabi-Yau} spaces through a process that generates singularities and is called conifold transitions \cite{Greene:1995hu}.
In this process, some fields become massless and then the spectrum of the fields is changed \cite{p}. Despite these interesting
properties, the general relativity is not well defined on singularities and sometimes it is necessary take off the conical
singularity. There are two main processes to smooth out the singularity: the first
one is called deformation because it deforms the quadric that defines the conifold; the second is called
resolution because it introduces a resolution parameter that controls the blow-up of singularity \cite{Candelas:1989js}.
These processes are used to study the extensions of AdS-CFT correspondence \cite{Klebanov:2000hb,Pando Zayas:2000sq}.

The change of spectrum on conifold spaces and the symmetry properties of the their smoothed versions have motivated
the study of those spaces in brane worlds cenarios. Firouzjahi and Tye \cite{Firouzjahi:2005qs} have studied the
behavior of the gravitational and Kaluza-Klein modes on deformed conifold and they have shown that the graviton has a
rather uniform probability distribution everywhere while a KK mode is peaked in the region near $r=0$. This region is called throat because it
has a big curvature and interpolates between asymptotically flat regions. Furthermore, Noguchi \textit{et al} \cite{Noguchi:2005ws}
have used the Klebanov-Strassler throat of a deformed conifold in order to obtain localized gravitational KK modes.
Since the supergravity solution of a 3-brane converges to $AdS_{5}\times S^{5}$ for $r\rightarrow 0$, Brummuer \textit{et al} \cite{Brummer:2005sh} have used a
throat
of conifold to deduce and extend the original Randall-Sundrum geometry. Further, V\'{a}zquez-Poritz \cite{VazquezPoritz:2001zt}
has shown that the $Z_{2}$ symmetry of the Randall-Sundrun model can be deduced from a dimensional reduction from a
six dimensional Eguchi-Hanson resolved conifold. Since this symmetry is natural in Eguchi-Hanson spaces, V\'{a}zquez-Poritz has
shown that the metric used for the localization of the gravity can de obtained from a particular conifold. Furthermore,
Pont\'{o}n and Poppitz \cite{Ponton:2000gi} have studied the relation between gravity localization on string-like defects and an AdS-CFT correspondence
on the so called hidden brane. Since the string-like geometry has a conical singularity far from the origin the authors have
found that the singularity could be resolved using the AdS-CFT duality. On the other hand, Kehagias \cite{Kehagias:2004fb}
had used a compact conical transverse space to explain the small value of the cosmological constant.
All of these points have motivated us to study geometries where the transverse space is a smoothed conifold and
ask whether that geometry could localize some kind of field in a 3-brane.

In this work we replaced the usual spherically symmetric transverse space by a 2-cycle of the resolved conifold.
Since the resolved conifold has spherical symmetry for a fixed $r$ and the radial metric component approaches to one
asymptotically, this geometry converges to string-like one if we put the 3-brane far from the tip of the cone. Another feature of the resolved conifold that has a great
importance
in the Randall-Sundrum-like model, is its
$\mathbb{Z}_{2}$ symmetry as pointed out in Ref. \cite{Pando Zayas:2000sq}. We have studied here the effects that variations of
the resolution parameter, or in other terms, the singularity, has on the localization of a scalar field in a 3-brane
placed in the origin of the resolved conifold. The study of effects of geometrical singularities has on the localization problems
has already done by Cvetic \textit{et al} \cite{Cvetic:2000dz} where the geometry was generated by a singular
domain wall as well by Gregory and Santos \cite{Gregory:2002tp} in the global vortex geometry. In present work, however, we
have chosen a transverse space whose singularity depends continually on a parameter.

The resolved conifold geometry also generalize the so called cigar-like geometries. Indeed, in cigar
manifolds the curvature is great but not infinity around the origin and flat asymptotically \cite{chow}. In
resolved conifold geometry the value of the
curvature in the origin is parameterized and asymptotically the curvature converges to zero or another constant.
Using a cigar-like geometry without cosmological constant Carlos and Moreno have found a supersymmetric solution that
has trapped gravity \cite{deCarlos:2003nq}. On the other hand, the so called Ricci flow is given by a parameter evolution of the metric through a heat-type equation
called Ricci equation \cite{chow,perelman,topping}.
This flow provides information about the stability of the manifold like the formation or blow up of singularities.
Therefore, it is interesting to use the smoothed conifold geometries to study the stability of
the bulk geometry in brane worlds using methods of geometric analysis like Ricci
equation.

This work is organized as follows: in the section \ref{Conifold} we have defined the metric for the resolved conifold
and we have studied its principal properties as well as its dependence on the resolution parameter. We also have defined
the conifold 2-cycle that we have worked out. In the section \ref{Bulk geometry} we have proposed a warped metric
ansatz, studied its Einstein equations
and compared it with the well studied string-like solutions. Still in this section, we have studied this geometry
for a linear and a nonlinear warp factor. In the section \ref{Scalarfieldinminimalcoupling}
we have studied the localization of a real scalar field
using a linear and a non-linear warp factor for both massive and massless modes. Finally we have concluded in the section \ref{Conclusions} summarizing our results and
presenting some perspectives.

\section{Conifold geometry}
\label{Conifold}
The 6-Conifold is a conical manifold $C_{6}\subset \mathbb{C}^{4}$ defined by the quadric \cite{Candelas:1989js}:

\begin{equation}
 z_{1}^{2}+z_{2}^{2}+z_{3}^{2}+z^{2}_{4}=0.
\end{equation}
The metric of a 6-conifold over a $X^{5}$ compact space is:
\begin{equation}
 ds^{2}_{6}=dr^{2}+r^{2}ds^{2}(X^{5}).
\end{equation}
This space has a naked singularity in $r=0$. For $X^{5}=T^{1,1}=SU(2)\times SU(2)/U(1)$ the metric is \cite{p,Klebanov:2000hb}:
\begin{eqnarray}
 ds^{2}_{6} & = & dr^{2}+\frac{r^{2}}{9}(d\psi+\cos\theta_{1}d\phi_{1}+\cos\theta_{2}d\phi_{2})^{2}\nonumber\\
            & + & \frac{r^{2}}{6}(d\theta_{1}^{2}+\sin^{2}\theta_{1} d\phi_{1}^{2}+d\theta_{2}^{2}+\sin^{2}\theta_{2} d\phi_{2}^{2}).
\end{eqnarray}
A smooth version of this conifold, called resolved conifold, has the metric \cite{Pando Zayas:2000sq,Cvetic:2000mh}

\begin{eqnarray}
 ds^{2}_{6} & = &
\left(\frac{r^{2}+6a^{2}}{r^{2}+9a^{2}}\right)dr^{2}+\frac{r^{2}}{9}\left(\frac{r^{2}+9a^{2}}{r^{2}+6a^{2}}\right)(d\psi+\cos\theta_{1}d\phi_{1}+\cos\theta_{2}d\phi_{2}
)^{2}\nonumber\\
            & + & \frac{1}{6}r^{2}(d\theta_{1}^{2}+\sin^{2}\theta_{1} d\phi_{1}^{2})+\frac{1}{6}(r^{2}+6a^{2})(d\theta_{2}^{2}+\sin^{2}\theta_{2} d\phi_{2}^{2}).
\end{eqnarray}

\begin{figure}
  \centering
 \includegraphics[scale=1.1]{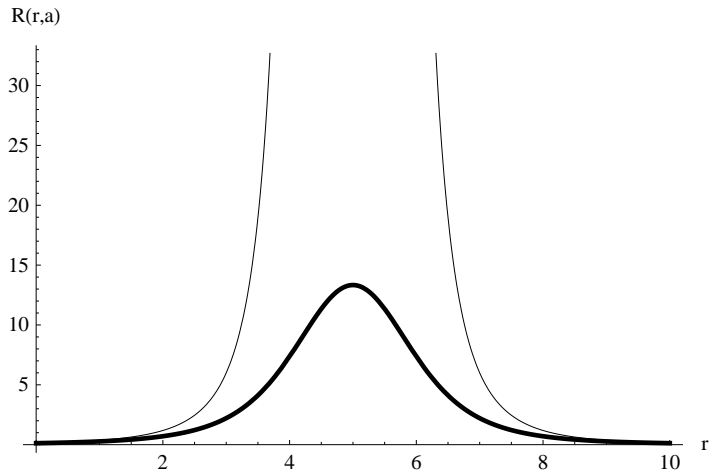}
 \caption{The scalar curvature of the resolved conifold. The origin was shifted to the point $r=5$. For $a=1$ (thick line), the function is regular in $r=0$
while for $a=0$ (thin line) the scalar curvature diverges in the origin.}
\label{resolvedscalarcurvature}
\end{figure}
We have plotted the scalar curvature in the figure (\ref{resolvedscalarcurvature}) where we have shifted the
origin to point $r=5$. Note that the curvature is smooth for $a=1$ in $r=0$
and the curvature diverges in the origin for $a=0$. Furthermore, the resolved conifold has a positive curvature and
is asymptotically flat. These issues have motivated us to use this manifold as a prototype of extension of transverse
spaces in the brane worlds. As a matter of fact, many authors have studied the localization of fields in spherical backgrounds
whose transverse space has positive, constant and non-singular curvature \cite{Cohen:1999ia,Gregory:1999gv,Gherghetta:2000qi,Oda:2000zc}.
Since the resolved conifold is
parameterized by the resolution parameter that controls the singularity at $r=0$, we can study the effects the singularity has on the
localization of fields in this particular space.

Note that in the limit $r\rightarrow 0$ the metric converges to a spherical one of radius $a$
\begin{equation}
 \lim_{r\rightarrow 0}{ds^{2}_{6}}=a^{2}(d\theta_{2}^{2}+\sin^{2}\theta_{2} d\phi_{2}^{2}),
\end{equation}
that has no singularity. Topologically this can be seen as a result of to take out a small neighborhood around $r=0$
and replaced it by a $S^{2}$ of radius $a$. Since in the limit $a\rightarrow 0$ we re-obtain the singular conifold again, the
radius $a$ can be used in order to measure how smooth is the conifold and then it is called resolution parameter.

Now if we take as constants our angular coordinates $\psi,\phi_{1},\theta_{2},\phi_{2}$, the cone 2-cycle can be written as the 2-resolved cone, namely
\begin{eqnarray}
 ds^{2}_{2} & = & \left(\frac{r^{2}+6a^{2}}{r^{2}+9a^{2}}\right)dr^{2}+ \frac{1}{6}(r^{2}+6a^{2})d\theta^{2}.
\end{eqnarray}

\begin{figure}
  \centering
  \includegraphics[scale=1.1]{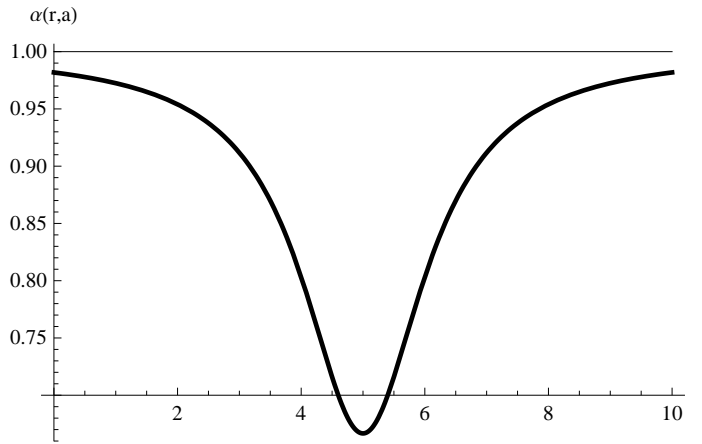}
\caption{Radial metric component of resolved conifold. For $a=1$ (thick line), the factor has a large variation and asymptotically
converges to one. For $a=0$ (thin line) radial factor
is equal to one as in the string-like defects geometry.}
 \label{resolvedconifoldmetric}
\end{figure}

This cone has a radial metric component $g_{rr}=\alpha(r)=\left(\frac{r^{2}+6a^{2}}{r^{2}+9a^{2}}\right)$ whose
graphic is plotted in the fig.(\ref{resolvedconifoldmetric}).
Note that $\lim_{r\rightarrow \infty}{g_{rr}}=1$ and therefore asymptotically the cone approaches the plane $\mathbb{R}^{2}$
with cylindrical metric of an effective radius $r_{eff}=\sqrt{\frac{(r^{2}+6a^{2})}{6}}$ which is the transverse metric
used in string-like geometries.
Near $r=0$ we have a hyperbolic behavior with high curvature and this region is called the throat. The angular
resolved conifold metric component $\beta(r,a)=\frac{(r^{2}+6a^{2})}{6}$ has a conical singularity dependent on
the resolution parameter. It is worthwhile to mention that as $a\rightarrow 0$ the width of the throat approaches zero.

Since the angular metric components diverge the effective radius of the base sphere grows without limit.
The scalar curvature of this 2-manifold is
 \begin{equation}
  R=R(r,a)=-\frac{6a^{2}(r^{2}+18a^{2})}{(r^{2}+6a^{2})^{3}}.
 \end{equation}
For sake of comparison we cite here the scalar curvature for the Hamilton cigar geometry \cite{chow}, namely
\begin{equation}
  R_{H}=\frac{4}{(1+r^{2})}.
 \end{equation}

Thus, this 2-cycle of the resolved conifold is a space of varying negative scalar curvature that converges asymptotically
to zero. This behavior is similar to the Hamilton cigar that is a Ricci soliton used in the study of the stability of manifolds \cite{topping}. The Hamilton
cigar is
an example of solution of the Ricci flow equation:
\begin{equation}
 \frac{\partial g_{ab}}{\partial t}= -2 R_{ab}
\end{equation}
where $t\in [0,1]$ is a parameter that describes the evolution of the geometry of the manifold. 


\section{Bulk geometry}
\label{Bulk geometry}
Once described the metric and the 2-cycle of the resolved conifold which we will use as a transverse space, we want
to study the localization of a scalar field in a 3-brane embedded in a six dimensional bulk of form $M_{6}=M_{4}\times \mathcal{C}_{2}$, where $\mathcal{C}_{2}$ is a
2-cycle of the resolved conifold described above.

The action for the gravitational is:

\begin{equation}
 S_{g} =\frac{1}{2K_{6}^{2}}\int_{\mathcal{M}_{6}}{d^{6}x\sqrt{-g}((R-2\Lambda)}+\int_{\mathcal{M}_{6}}{dx^{6}\sqrt{-g}L_{m}}
\end{equation}
where $K_{6}^{2}=\frac{8\pi}{M_{6}^{4}}$ and $M_{6}^{4}$ is the six-dimensional bulk Planck mass.
Further, let us assume the following ansatz for the energy-momentum tensor:

\begin{align}
 T^{\mu}_{\nu} & = t_{0}(r)\delta^{\mu}_{\nu}\\
 T^{r}_{r} & = t_{r}(r)\\
 T^{\theta}_{\theta} & = t_{\theta}(r)
\end{align}
where,

\begin{equation}
 T_{ab}=-\frac{2}{\sqrt{-g}}\frac{\partial L_{m}}{\partial g^{ab}}
\end{equation}\\

Now, let us choose the metric components in such way that we can obtain the string-like defect geometry in some limits.
For now on, the metric ansatz will be
\begin{eqnarray}
 \label{ansatz}
 ds^{2}_{6} & =  & e^{-A(r)}\hat{g}_{\mu\nu}dx^{\mu}dx^{\nu}+\alpha(r,a) dr^{2} + \beta(r,a) e^{-B(r)}d\theta^{2},
\end{eqnarray}
where $a\geq0$ is the resolution parameter. This ansatz extends the solution for string-like defects
by the inclusion of the resolved conifold metric factors $\alpha(r,a), \beta(r,a)$ . Since the radial component
approaches to one at infinity,  if we put the 3-brane in a point far from the origin of the resolved conifold,
this ansatz goes to the Oda ansatz one \cite{Oda:2000zc}. Furthermore, the geometry of the bulk is parameter-dependent which
enable us to control the singularity.

%

The Einstein equations for the metric ansatz in eq. (\ref{ansatz}) are
\begin{align}
 3A''+B''-\frac{3}{2}A'B'-B'^{2}-3A'^{2}
+\frac{3}{2}\left(\frac{\beta'}{\beta}-\frac{\alpha'}{\alpha}\right)A'+\nonumber\\
\left(\frac{\beta'}{\beta}-\frac{1}{2}\frac{\alpha'}{\alpha}\right)B'-\frac{\beta''}{\beta} \label{eq3.1}
+\frac{1}{2}\frac{\alpha'}{\alpha}\frac{\beta'}{\beta}+\alpha\left(\frac{1}{2}e^{A}\hat{R} -2\Lambda+2K_{6}^{2}t_{0}\right) & =  0,\\
-3A'^{2}-2A'B'+2\frac{\beta'}{\beta}A'+\alpha\left(e^{A}\hat{R}-2\Lambda+2K_{6}^{2}t_{r}\right) & =  0,\label{eq3.2} \\
 4A''-5A'^{2}-2\frac{\alpha'}{\alpha}A'+\alpha\left(e^{A}\hat{R} -2\Lambda + 2K^{2}_{6}t_{\theta}\right)& =  0.\label{eq3.3}
\end{align}
The continuity equation for the energy-momentum tensor is
\begin{equation}
 \nabla^{a}T_{ab}=0.
\end{equation}
This equation yields a constraint on the components of energy-momentum tensor
\begin{equation}
 t'_{r}=2A'(t_{r}-t_{0})+\frac{B'}{2}(t_{r}-t_{\theta})+\frac{\beta'}{2\beta}(t_{\theta}-t_{r}). \label{continuity}
\end{equation}

The equations (\ref{eq3.1}, \ref{eq3.2}, \ref{eq3.3}) and the continuity equation (\ref{continuity}) differ from the solution of the string-like defects by the addition
of the angular factor $\beta$ of the resolved conifold metric.

Let us now sum the radial and angular Einstein equations and assume that $A(r)=B(r)$. This yields a linear differential
equation for $A(r)$ in the form
\begin{equation}
\label{Aequation}
 2A''(r)-\left(\frac{\alpha'}{\alpha}+\frac{\beta'}{\beta}\right)A'(r)+K_{6}^{2}\alpha(t_{\theta}-t_{r})=0.
\end{equation}
Defining
\begin{eqnarray}
 \delta(r,a) & = & -\left(\frac{\alpha'}{\alpha}+\frac{\beta'}{\beta}\right),\\
 \chi(r,a)   & = & K_{6}^{2}\alpha(t_{\theta}-t_{r}),
\end{eqnarray}
the eq. (\ref{Aequation}) can be rewritten as
\begin{equation}
\label{A'equation}
 2A''(r)+\delta(r,a)A'(r)+\chi(r,a)=0.
\end{equation}
The solution of eq.(\ref{A'equation}) is

\begin{equation}
 A(r)= A(0)-\int_{0}^{r}{\left(\frac{\int_{0}^{r'}{\eta(r'',a)\chi(r'',a) dr''}}{\int_{0}^{r'}{\eta(r'',a)dr''}}\right)dr'.	}
\end{equation}
Let us suppose the boundary conditions
\begin{equation}
\label{boundary1}
 A(0) =  q,
\end{equation}
\begin{equation}
\label{boundary2}
 \lim_{r\rightarrow \infty}{A(r)} = \infty,
\end{equation}
where $q$ is a constant.

The equation (\ref{A'equation}) with boundary conditions gives the warp factor. In the point $r=0$, the metric
defined by eq. (\ref{ansatz}) becomes
\begin{eqnarray}
 \label{fiber}
 ds^{2}_{6} & =  & \hat{g}_{\mu\nu}dx^{\mu}dx^{\nu}+ \frac{a^{2}}{6}d\theta^{2}.
\end{eqnarray}
This is a factorizable metric of the space $M_{4}\times S^{1}$ where $S^{1}$ has radius $\frac{a}{\sqrt{6}}$.
Therefore, the 3-brane can be realized as a normal fiber bundle of strings in $r=0$.

In this geometry, the relationship between the four-dimensional Planck mass ($M_{4}$) and the bulk Planck mass
($M_{6}$) is given by

\begin{equation}
  \label{planckmass}
 M^{2}_{4}=2\pi M_{6}^{4}\int_{0}^{\infty}{e^{-A(r)-\frac{B(r)}{2}}\sqrt{\alpha(r,a)\beta(r,a)}dr}.
\end{equation}
Therefore, we can use the resolution parameter in order to tuning the ratio between the Planck masses and so explaining the
hierarchy between them. This is an extension of the string-like tuning of the Planck masses: in the string-like
geometry, the adjust is made by the six-dimensional cosmological constant $\Lambda$ and the tension of the string $\mu$
\cite{Gherghetta:2000qi,Oda:2000zc}. Here, we have added a dependence on a geometrical parameter $a$. Note that the hierarchy is well-defined even for
the singular cone $(a=0)$. Therefore, using parameter dependent transverse spaces we could obtain a parameter
dependent hierarchy. Since there are many parameter-dependent spaces these manifolds could be used to solve the
hierarchy problem. We argue that this dependence could be related to possible transformations in transverse space,
for instance, the conical transitions.

\subsection{Linear warp factor}
\label{linear warp factor}

Now let us choose a specific warp factor $A(r)$ and study his geometrical consequences.
Let us choose the linear warp factor, i.e., $A(r)=kr$, where $k$ is a real constant. This warp factor was widely used
both in Randall-Sundrum models \cite{Randall:1999vf,Randall:1999ee} and in string-like geometries \cite{Gregory:1999gv,Gherghetta:2000qi,Oda:2000zc}. This warp factor
was the first used to solve the hierarchy problem. Further, still in Randall-Sundrum and string-like geometries, it provides a $AdS_{6}$ geometry to the
bulk, i.e., a maximally symmetric space with negative cosmological
constant.

With that choice the Einstein equations become
\begin{align}
B''-\frac{3k}{2}B'-B'^{2}-3k^{2}
+\frac{3k}{2}\left(\frac{\beta'}{\beta}-\frac{\alpha'}{\alpha}\right)+\left(\frac{\beta'}{\beta}-\frac{1}{2}\frac{\alpha'}{\alpha}\right)B'\nonumber\\
-\frac{\beta''}{\beta}
+\frac{1}{2}\frac{\alpha'}{\alpha}\frac{\beta'}{\beta}+\alpha\left(\frac{1}{2}e^{A}\hat{R} -2\Lambda+2K_{6}^{2}t_{0}\right) & =  0,\\
-3k^{2}-2kB'+2\frac{\beta'}{\beta}k+\alpha\left(e^{A}\hat{R}-2\Lambda+2K_{6}^{2}t_{r}\right) & =  0,\\
-5k^{2}-2\frac{\alpha'}{\alpha}k+\alpha\left(e^{A}\hat{R} -2\Lambda + 2K^{2}_{6}t_{\theta}\right)& =  0.\label{tteta}
\end{align}
Summing the radial and angular equations above, we obtain the solution for the warp factor $B(r)$:
\begin{equation}
 B(r)=kr+\ln(\alpha(r,a) \beta(r,a))+\frac{K_{6}^{2}}{k}\int_{0}^{r}{\alpha(r',a)(t_{r}(r')-t_{\theta}(r'))dr'}.
\end{equation}
Therefore, we can get $B(r)=A(r)$ if
\begin{equation}
\label{A=B}
 \ln(\alpha(r,a) \beta(r,a))+\frac{K_{6}^{2}}{k}\int_{0}^{r}{\alpha(r',a)(t_{r}(r')-t_{\theta}(r'))dr'}=0.
\end{equation}
Equation (\ref{A=B}) provides a constraint between the resolved conical geometry and the content of matter.
In vacuum, the warp factor $B(r)$ is given by
\begin{equation}
 \label{Bvaccum}
B(r)=kr+\ln(\alpha(r,a) \beta(r,a))
\end{equation}

The solution for the function $B(r)$ above differs from the string-like defect one by the conifold
metric components $\alpha,\beta$ \cite{Oda:2000zc}. Besides, note that in general $B(r)$ depends on the resolution
parameter. Therefore, we can make $\beta(r,a)=1$ and still detects the effects of the resolution on the conifold.

Let us suppose now an angular energy-momentum tensor of the form

\begin{equation}
 t_{\theta}(r)= \zeta e^{kr}+\lambda(r)+\rho,
\end{equation}
where $\zeta,\rho$ are constants and
\begin{equation}
 \lambda(r)=\frac{1}{4K_{6}^{2}\alpha(r)}\left(5P^{2}+2P\frac{\alpha'}{\alpha}-3\left(\frac{\alpha'}{\alpha}\right)^{2}\right).
\end{equation}
Now let us suppose the 3-brane $M_{4}$ is a maximally symmetric space. Therefore, we can define a
3-cosmological constant $\Lambda_{3}$, satisfying
\begin{equation}
 \hat{R}_{\mu\nu}-\frac{\hat{R}}{2}\hat{g}_{\mu\nu}=-\Lambda_{3}\hat{g}_{\mu\nu}
\end{equation}
Thus, its scalar curvature $\hat{R}$ must be
constant. Therefore, from equation (\ref{tteta}) we conclude that

\begin{eqnarray}
 \zeta & = & \frac{\hat{R}}{2K_{6}^{2}},\\
 \rho & = & \frac{\Lambda}{K_{6}^{2}},\\
  k & =& P.
\end{eqnarray}
It is worthwhile to mention that the solution above for $t_{\theta}$ differs of string-like type by the terms $\alpha$
and $\frac{\alpha'}{\alpha}$ and so we obtain the string-like defect as a special case of the resolved conifold one.

For $B(r)=A(r)$ the components of the energy-momentum tensor are
\begin{eqnarray}
 t_{r}(r) & = & \frac{\Lambda}{K_{6}^{2}}-\frac{e^{kr}\hat{R}}{K_{6}^{2}}+\frac{1}{\alpha K_{6}^{2}}(5k^{2}-2\frac{\beta'}{\beta}k),\\
 t_{0}(r) & = & \frac{\Lambda}{K_{6}^{2}}-\frac{e^{kr}\hat{R}}{2K_{6}^{2}}+\frac{1}{2\alpha K_{6}^{2}}\left(\frac{11}{2}k^{2},
+2(\frac{\beta'}{\beta}-\frac{\alpha'}{\alpha})k-\frac{\beta''}{\beta}+\frac{1}{2}\frac{\alpha'}{\alpha}\frac{\beta'}{\beta}\right).
\end{eqnarray}
Since $\beta'$ diverges, then for $\hat{R}\leq 0$ the component $t_{0}$ satisfies the energy dominant condition,
$t_{0}\geq 0$.

This linear warp factor satisfies the boundary conditions (\ref{boundary1}) and (\ref{boundary2}). Since the warp factor diverges asymptotically we can impose the
following condition
\begin{equation}
 A'(r)>0.
\end{equation}
Now let us analyze the asymptotic behavior of the linear warp factor. The angular Einstein equation is

\begin{equation}
 -5k^{2}-2\frac{\alpha'}{\alpha}k+\alpha\left(e^{A}\hat{R} -2\Lambda + 2K^{2}_{6}t_{\theta}\right)=  0
\end{equation}
Asymptotically
$\lim_{r\rightarrow\infty}{\frac{\alpha'}{\alpha}}=0$ and $\lim_{r\rightarrow\infty}{\alpha=1}$.
Thus, for $t_{\theta}=-\frac{e^{kr}\hat{R}}{2K_{6}^{2}}$, the angular equation becomes:
\begin{eqnarray}
 5k^{2}+2\Lambda=0 & \Rightarrow & \Lambda<0
\end{eqnarray}
Therefore, the bulk is asymptotically $AdS_{6}$ for the linear warp factor. Further, since
\begin{equation}
 R=3\Lambda -\frac{K_{6}^{2}}{2}T
\end{equation}
the energy-momentum tensor has a core around the brane which is very similar to the string-like geometry
\cite{Gherghetta:2000qi}.

\begin{figure}
 \centering
 \includegraphics[scale=1.1 ,bb=-41 -80 233 146]{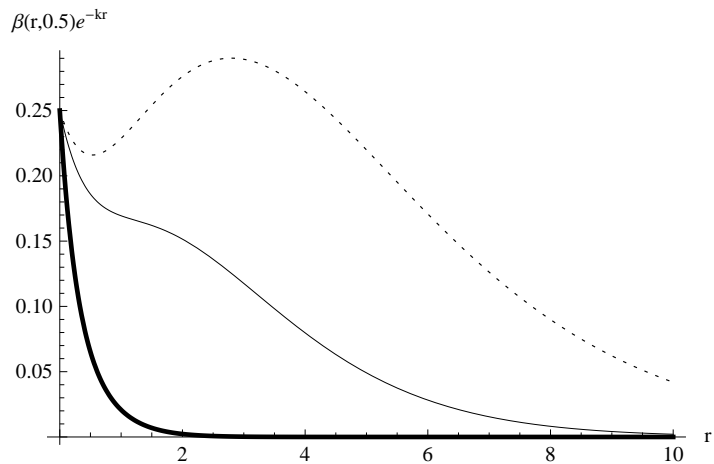}
\caption{Angular metric factor of bulk for $a=0.5$. For $k=3$, we have a monotonic exponential behavior characteristic
of the string-like defects and cigar geometries (thick line).
For $k=0.9$ (thin line) and $k=0.6$ (dotted line) the angular factor grows until a maximum and then decreases
exponentially. Note that the former behavior makes the angular component vanish more slowly.}
 \label{fatormetricoangular}
\end{figure}

\begin{figure}
\centering
 \includegraphics[scale=1.1, bb=-41 -80 233 146]{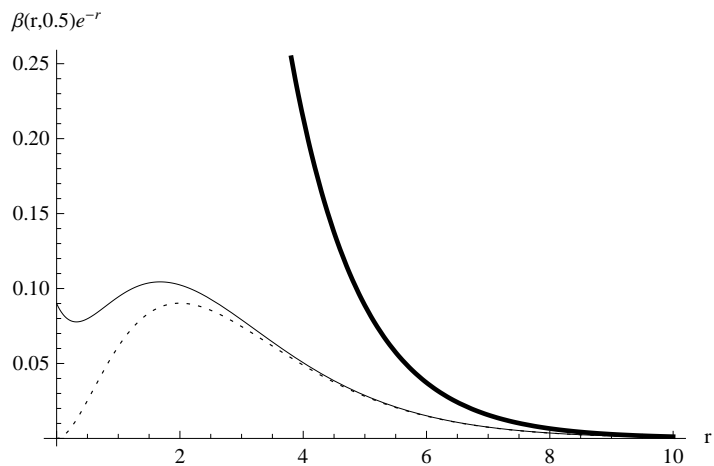}
\caption{Angular metric component for $k=1$. For $a=3$ the component decreases exponentially like for string-like
defects and cigar geometries (thick line). For $ a \simeq 0.3 $ (thin line) there is no conical singularity and the component reaches a maximum before
decay. For $a=0$ (dotted line) the component begin with a conical singularity at the origin, increases until a maximum and then vanish slowly.}
\label{f}
\end{figure}

Another important feature in this approach is the freedom on the factor $k$ since the constant $P$ must only to be positive.
We have plotted the angular component for some different values of $k$  and for $a=0.5$ in
fig.(\ref{fatormetricoangular}). For $k=3$, this component has an exponential decreasing behavior and then approaches to the
configuration of string-like defects and cigar geometries \cite{Gregory:1999gv,Gherghetta:2000qi,Oda:2000zc,deCarlos:2003nq}.
However, for $0.5\leq k \leq 1$ the angular component increases until reaches a maximum and then
decreases exponentially. Hence, the decreasing of the $k-$parameter has the feature of
damping the exponential decreasing of the angular component.
The same features appear if we fix $k$ and vary the resolution parameter. Indeed, as $a\rightarrow 0$ a peak arises
making the angular component vanish more slow as shown in the figure (\ref{f}). Besides, the angular metric component
has the same asymptotic behavior for any value of $a$ but it changes its behavior close to the brane.
%

\begin{figure}
\centering
 \includegraphics[scale=1.1, bb=-41 -80 233 146]{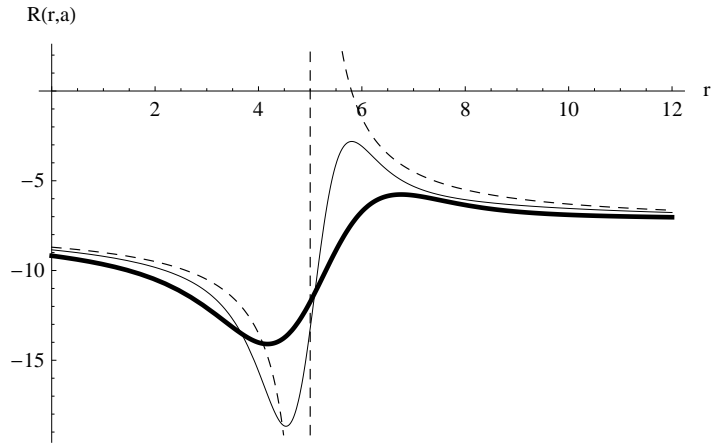}
\caption {Bulk scalar curvature for the linear warp factor $(k=1)$. We have put the brane in $r=5$. For $a=1$ (thick line) the curvature increases until
an asymptotic value around $-7.5$. For $a=0.5$ (thin line) the curvature grows until reach a maximum and then decreases to the
same asymptotic value. For $a=0$ (dashed line) the curvature diverges on the brane.}
\label{linearscalarcurvature}
\end{figure}

For the linear warp factor the scalar curvature is given by


\begin{equation}
 R=-\frac{3}{2}\frac{(1620a^{6}k^{2}+kr^{5}(5kr-4)+12a^{4}(60k^{2}r^{2}-13kr+6)+a^{2}r^{2}(105k^{2}r^{2}-50kr+4))}{(r^{2}+6a^{2})^{3}}
\end{equation}
whose graphic is plotted in the fig.(\ref{linearscalarcurvature}).
For better viewing, let us make the change of variable
\begin{equation}
r\rightarrow r-5.
\end{equation}
Therefore, the bulk has a varying negative scalar curvature.

%

\subsection{Nonlinear warp factor}
\label{Nonlinear warp factor}
\begin{figure}
 \centering
 \includegraphics[scale=1.1,bb=-41 -80 233 146]{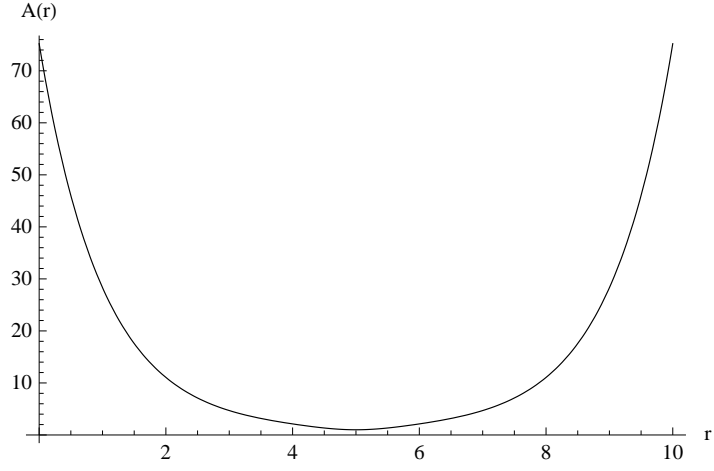}
\caption{Nonlinear warp factor. This function is symmetric around the brane and diverges asymptotically.}
\label{nonlinearwarpfactor}
\end{figure}

\begin{figure}
 \centering
 \includegraphics[scale=1.1,bb=-41 -80 233 146]{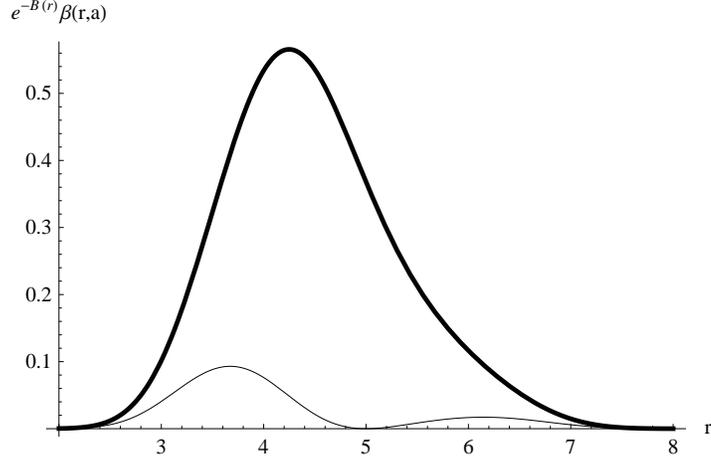}
\caption{Bulk angular metric factor for the nonlinear warp factor. For $a=1$ (thick line) the component is similar
to a gaussian function and is not zero on the brane - there is no conical singularity. For $a=0$ the component
has two maxima and it has a conical singularity.}
\label{warpedangularcomponent}
\end{figure}

In addition to the configuration above, we have used another warp factor slightly different from the warp factor previously studied
by Fu \textit{et al} \cite{Fu:2011pu}. Our proposed warp factor is given by

\begin{equation}
 A(r)=B(r)=\cosh(r)+\tanh(r)^{2}.
\end{equation}
Note that, like in Randall-Sundrum model where the warp factor is a modulus function, this nonlinear warp factor is
symmetric with respect to reflection on the brane, i.e., it has $Z_{2}$ symmetry, as shown in the figure (\ref{nonlinearwarpfactor}). Furthermore, it gives a localized
angular component as seen in
the fig. (\ref{warpedangularcomponent}).
We have plotted the scalar curvature for this warp factor for $a=0$ and $a=1$. Note that the behavior of the scalar
curvature
is opposite of the linear case because for the non-linear case the scalar curvature
is regular for $r=0$ but diverges at infinity. Moreover, the curvature is positive around the origin and is
negative for large distances.

Therefore, the geometry for the warp factor $A(r)=cosh(r)+tanh(r)^{2}$ has a well-behavior in the origin but diverges
asymptotically. Furthermore, asymptotically the warp factor satisfies
\begin{eqnarray}
 lim_{r\rightarrow\infty}A'(r)=0,\\
 lim_{r\rightarrow\infty}A''(r)>0.
\end{eqnarray}
Then, for $t_{\theta}=0$ and $\hat{R}=0$ the angular Einstein equation satisfies
\begin{eqnarray}
 4A''(r)-2\Lambda=0 & \Rightarrow \Lambda>0.
\end{eqnarray}
However, since the scalar curvature diverges far from the brane, we can not conclude that the bulk converges to
the $dS_{6}$ space. Indeed, since the scalar curvature diverges at infinite and
\begin{eqnarray}
 R=K_{6}^{2}T+6\Lambda &\Rightarrow& \lim_{r\rightarrow \infty}{T}=-\infty,
\end{eqnarray}
whatever produces this
geometry, the scalar curvature has a radial energy-momentum tensor component that diverges asymptotically. This feature contrasts a lot
with the string-like geometry.
\begin{figure}
 \centering
 \includegraphics[scale=1.1,bb=-41 -80 233 146]{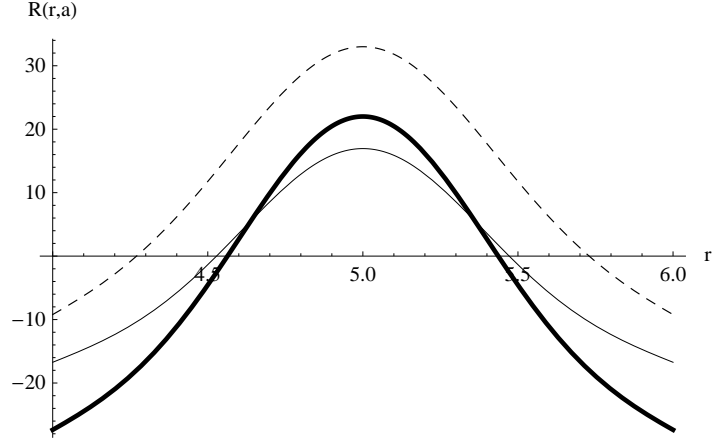}
\caption{Bulk scalar curvature for the nonlinear warp factor.}
\label{curvaturanaolinear}
\end{figure}
\section{Scalar field in minimal coupling}
\label{Scalarfieldinminimalcoupling}
Now let us study the localization of a scalar fields in the geometry analyzed so far.
The action for scalar field minimally coupled to the gravity is
\begin{equation}
 S_{s}=\int_{\mathcal{M}_{6}}{dx^{6}\sqrt{-g}g^{ab}\partial_{a}\phi \partial_{b}\phi}
\end{equation}
The equation of motion for the scalar
field is given by

\begin{equation}
 \partial_{A}(\sqrt{-g}g^{AB}\partial_{B}\Phi)=0
\end{equation}
Let us assume that this scalar field is a product of a 4-component field with Poincar\'{e} symmetry and another scalar field
living only in the 2-cycle of conifold, i.e.,

\begin{equation}
 \Phi(x^{\mu},r,\theta)=\hat{\phi}(x^{\mu})\tilde{\phi}(r,\theta).
\end{equation}
From Poincar\'{e} symmetry the 3-brane scalar field must satisfies the mass condition
\begin{equation}
 \partial_{\mu}\partial^{\mu}(\hat{\phi}(x^{\mu}))=m^{2}\hat{\phi}(x^{\mu}).
\end{equation}
Since $0\leq \theta \leq 2\pi$, let us assume that $\tilde{\phi}(r,\theta)$ can be expanded in Fourier series as
\begin{equation}
\label{transverseansatz}
 \tilde{\phi}(r,\theta)=\chi(r)\sum_{l=0}^{\infty}{e^{il\theta}}.
\end{equation}
Using the ansatz $(\ref{transverseansatz})$ yields
\begin{equation}
\label{stequation1}
 \left(\frac{\sqrt{-g}}{\alpha(r,a)}\chi'(r)\right)'-\frac{l^{2}\sqrt{-g}e^{B}}{\beta(r,a)}\chi(r) + m^{2}\sqrt{-g}e^{A}\chi(r)=0
\end{equation}
Equation $(\ref{stequation1})$ is a Sturm-Liouville like equation. Further, let us looking for solutions that satisfy the boundary conditions
\begin{equation}
\chi'(0)= \lim_{r\rightarrow\infty}\chi'(r)=0.
\end{equation}
If we have two solutions of eq. $(\ref{stequation1})$, namely $\chi_{i}(r)$ and $\chi_{j}(r)$, the
orthogonality
relations between them are
\begin{equation}
 \int_{0}^{\infty}{\sqrt{\alpha(r,a)\beta(r,a)}e^{(-A(r)-\frac{B(r)}{2})}\chi_{i}*\chi_{j}dr}=\delta_{ij}
\end{equation}
We can rewrite eq. (\ref{stequation1}) as 
\begin{equation}
\label{radialequation}
 \chi''(r)-\frac{1}{2}\left(4A'+B'+\frac{\alpha'}{\alpha}-\frac{\beta'}{\beta}\right)\chi'(r)+
\alpha e^{A}\left(m^{2}-l^{2}\frac{e^{B-A}}{\beta}\right)\chi(r)=0.
\end{equation}

Note that equation (\ref{radialequation}) is similar to that
found in string-like geometries \cite{Gherghetta:2000qi,Oda:2000zc}, regardless the conifold terms
$\alpha(r,a),\beta(r,a)$.
Further, we can see from equation (\ref{radialequation}) that it is possible to add the conifold terms to the warp
factors. Here we can choose two distinct paths. On one hand it is possible to study the resolution behavior directly in the warp factors. On other hand, we
can study the localization in a factorized geometry, i.e., without the exponential warp factors $e^{A(r)}, e^{B(r)}$.
Besides, we could define an effective angular
number $l_{eff}=\frac{l^{2}}{\beta(r,a)}$ which would depends on the point and on the resolution parameter.

\subsection{Massive modes}
\label{Massive modes}
Let us simplify the equation (\ref{radialequation}) making the following change of variable
\begin{equation}
\label{change}
 z=z(r)=\int_{0}^{r}{\alpha(r')^{\frac{1}{2}}e^{\frac{A(r')}{2}}dr'}.
\end{equation}
Since the radial metric component $\alpha(r,a)$ is a non-negative smooth function of $r$, i.e.,
$\forall r\in [0,\infty), \alpha(r,a)>0$, so for a fixed $a$, $\alpha(r_{1})\neq \alpha(r_{2})\Leftrightarrow r_{1}\neq r_{2}. $
Thus,
\begin{equation}
 \frac{dz}{dr}>0,
\end{equation}
and so $z(r)$ is a smooth, monotonic increasing function of $r$.

Using the change of variable in eq.(\ref{change}) the equation (\ref{radialequation}) turns to be
\begin{equation}
 \ddot{\chi}(z)-\frac{1}{2}\left(3\dot{A}+\dot{B}-\frac{\dot{\beta}}{\beta}\right)\dot{\chi}(z)+
\left(m^{2}-l^{2}\beta^{-1}e^{B-A}\right)\chi(z)=0.
\label{motionequation}
\end{equation}
In order to simplify further, let us write $\chi(z)$ in the form
\begin{equation}
 \chi(z)=e^{\frac{(3A+B-\ln(\beta))}{4}}\Psi(z).
\end{equation}
From equation (\ref{motionequation}), the $\Psi(z)$ function must obey
\begin{eqnarray}
 -\ddot{\Psi}(z)+V(z)\Psi(z)=m^{2}\Psi(z),
\end{eqnarray}
where
\begin{eqnarray}
V(z) & = & \left(\frac{3\dot{A}+\dot{B}-\beta^{-1}\dot{\beta}}{4}\right)^{2}-\frac{\left(3\ddot{A}+\ddot{B}+
\beta^{-2}(\dot{\beta})^{2}-\beta^{-1}\ddot{\beta}\right)}{4}\nonumber\\
     & + & l^{2}\beta^{-1}e^{B-A}
\end{eqnarray}
This is a time-independent Schroedinger-like equation. We can study the localization of the scalar field by analyzing the behavior of the
potential around a potential well. Returning to $r$ coordinate the potential can be written as
\begin{eqnarray}
 V(r,a,l) & = & \frac{e^{-A}}{\alpha}\Big\{\frac{1}{16}(15A'^{2}+B'^{2}+8A'B')-\frac{1}{4}\left(3A''+B''\right)\nonumber\\
      & + & (3A'+B')\frac{\alpha'}{8\alpha}-\frac{\alpha'\beta'}{8\alpha\beta}
     -\frac{\beta'}{8\beta}A'+\left(\frac{\beta'}{4\beta}\right)' \Big\}+l^{2}\beta^{-1}e^{B-A}
\end{eqnarray}
The study of this potential provides graphic information about the possible stable solutions around a minimum.
\subsubsection{Linear warp factor}
\begin{figure}
 \centering
  \includegraphics[scale=1.1,bb=0 0 240 161]{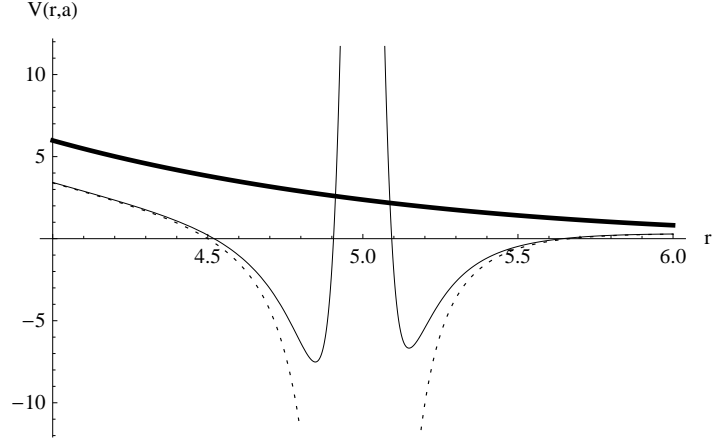}
\caption{Potential for linear warp factor and $l=0$. For $a=1$ (thick line) the potential
decays exponentially. For $0.1< a <0$ (dashed line) it appears two asymmetric potential wells beyond the point
$r=5$ but
in this point the potential is a maximum. For $a=0$ a potential well arises on the brane (dotted line)}.
\label{potenciallinear1}
\end{figure}

Now, let us analyze the case for the linear warp factor keeping the condition $A(r,k)=B(r,k)$.
The potential is given by
\begin{eqnarray}
 V(u,a,l,k) & = &
\frac{e^{-ku}}{\alpha}\Big\{\frac{3}{2}k^{2}+\frac{k}{2}\frac{\alpha'}{\alpha}-\frac{1}{8}k\frac{\beta'}{\beta}-
\frac{1}{8}\frac{\alpha'}{\alpha}\frac{\beta'}{\beta}\nonumber\\
            & + & \frac{1}{4}\left(\frac{\beta'}{\beta}\right)'\Big\}+l^{2}\beta^{-1}
\end{eqnarray}
We have plotted the potential for $l=0,k=1$ in the figure (\ref{potenciallinear1}). For large values of $a$
the potential well decay exponentially. As $a\rightarrow 0$ two asymmetric minima arise far from the brane.
However, for $a=0$, the point $r=5$ turns to be an infinite potential well. Therefore, it is possible to find
a localized solution of eq. (\ref{motionequation}). Indeed, the eigenfunction $\chi(r)$ must satisfies
the differential equation
\begin{equation}
\label{linearradialequation}
 \chi''(r)+\left(\frac{(r-5)}{(r-5)^{2}+9a^{2}}-\frac{5}{2}k\right)\chi'(r)+\left(\frac{(r-5)^{2}+6a^{2}}{(r-5)^{2}+9a^{2}}\right)e^{k(r-5)}m^{2}\chi(r)=0.
\end{equation}
For $\left(\frac{a}{r-5}\right)\rightarrow 0$, which is valid for the singular cone ($a=0$) and for distant points
($(r-5)\rightarrow \infty$),
the radial function satisfies
\begin{equation}
\label{massive1}
 \chi''(r)+\left(\frac{1}{(r-5)}-\frac{5}{2}k\right)\chi'(r)+e^{k(r-5)}m^{2}\chi(r)=0.
\end{equation}
Note that eq. (\ref{massive1}) differs from the string-like \cite{Gherghetta:2000qi} equation for massive modes by the factor $\frac{1}{r-5}$. For
$(r-5)\gg\max(a,\frac{2}{2k})$, eq.(\ref{massive1}) converges to well-known equation of string-like defects \textbf{\cite{Gherghetta:2000qi}}
\begin{equation}
\label{massivegs}
 \chi''(r)-\frac{5}{2}k\chi'(r)+e^{kr}m^{2}\chi(r)=0,
\end{equation}
and so asymptotically the field has
the same features as in string-like geometry. For instance, the spectrum of mass has the same asymptotic behavior.


In order to study the behavior of massive modes near the brane let us take the limit $\left(\frac{r-5}{a}\right)\rightarrow 0$, $a\neq 0$ and therefore the equation
for the
radial component becomes
\begin{equation}
\label{massive2}
 \chi''(r)+\left(\frac{r-5}{9a^{2}}-\frac{5}{2}k\right)\chi'(r)+\frac{2}{3}e^{k(r-5)}\chi(r)=0.
\end{equation}
We can also use eq. (\ref{massive2}) to study the asymptotic
behavior of the eingenfunction for singular conifold $a=0$ whereas $\left(\frac{r-5}{a}\right)\rightarrow 0$.

For $|r-5|\ll \frac{45a^{2}k}{2}$, eq. (\ref{massive2}) turns to be the string-like equation for massive modes \cite{Gherghetta:2000qi}. Hence, the resolved
conifold geometry resembles asymptotically and close to the brane the string-like geometry. Therefore, for $(r-5)\rightarrow 0$ or $(r-5)\rightarrow \infty$ we have
the well-known string-like solution
\begin{equation}
\label{assolution}
 \chi(r)\rightarrow
e^{\frac{5}{4}cr}\left(C_{1}J_{\frac{5}{2}}\left(\frac{2m}{c}e^{\frac{c}{2}}r\right)+C_{2}Y_{\frac{5}{2}}\left(\frac{2m}{c}e^{\frac{c}{2}}r\right)\right)
\end{equation}
However, this solution in eq. ($\ref{assolution}$) is not normalizable for $[0,\infty)$. In order to normalize the field we can use a cut-off distance $r_{c}$
and apply the boundary condition in the point $r=r_{c}$ instead of the infinity. Then we take the limit $r_{c}\rightarrow \infty$ and analyze the behavior of the mass
spectrum. Since the solution of eq. $(\ref{linearradialequation})$ behaves near and far the brane like the Bessel functions, for $r_{c}>\max(a,\frac{2}{2k})$, the
asymptotic mass spectrum must be of the form
\begin{equation}
 m_{n}=c\left(n-\frac{1}{2}\right)\frac{\pi}{2} e^{-\frac{cr_{c}}{2}}
\end{equation}

\begin{figure}
 \centering
  \includegraphics[scale=1.1,bb=0 0 240 161]{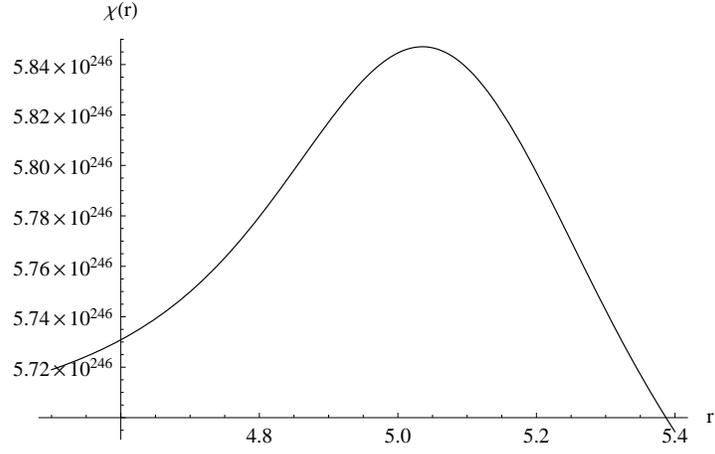}
\caption{Eigenfunction for linear warp factor and $a=1,k=1$. The solution is defined for points closed to the brane.
The field has compact support
and is well defined on the brane $(r=5)$. }
\label{massivemode1}
\end{figure}

On the other hand, expanding the exponential until first order yields
\begin{equation}
 \chi''(r)+\left(\frac{r-5}{9a^{2}}-\frac{5}{2}k\right)\chi'(r)+\frac{2}{3}(1+k(r-5))\chi(r)=0.
\end{equation}
Therefore, this equation describes either the behavior of the field in the neighborhood of the brane or the asymptotic
behavior of the eigenfunction for $a\neq 0$ . Their solution is the
product between an exponential function and the confluent hypergeometric function of second kind, namely
\begin{equation}
 \chi(r,a)=E(r,a)M\left(\frac{(1 - 6 a^{2} - 135 a^{4} - 324 a^{6})}{2}
, \frac{1}{2},-\frac{10+45a^{2}+216a^{4}-2r}{6\sqrt{2}a}\right),
\end{equation}
where, $E(r,a)=e^{\frac{(10+45a^{2}+108a^{4}-r)}{18a^{2}}}$. The graphic of this function was plotted in the fig. (\ref{massivemode1}).

\subsubsection{Nonlinear warp factor}
\begin{figure}
 \centering
 \includegraphics[scale=1.1,bb=0 0 240 157]{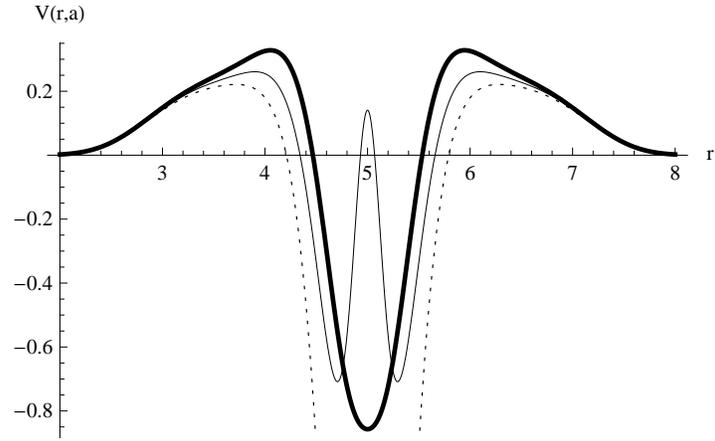}
  \caption{Potential for the nonlinear warp factor. For $a=0.24$ (thick line) there is only one potential well around
the brane (the usual \textit{volcano} potential); for $a=0.16$, there is a maximum in the brane and the formation of two minima besides the brane
(thin line); as $a\rightarrow 0$ the width of the maximum decreases and for $a=0$ the potential has a infinite minimum on
the brane (dashed line).}
\label{potencialnaolinearl0}
\end{figure}

Now let us study the solutions for the nonlinear warp factor. First, we have plotted the potential using several values of the resolution parameter $a$ and
$l=0$ in the figure (\ref{potencialnaolinearl0}). For $a\geq 0.5$ there is a potential well on the brane ($r=5$) and
then there are massive modes trapped to the brane. Nevertheless, for $0.5<a< 0$, there is a potential barrier on
the brane and there is only a potential well beside the brane. However, for $a=0$ a
potential well appears again on the brane and thus there are localized states on the brane in the singular
conifold.

\begin{figure}
 \centering
 \includegraphics[scale=1.1,bb=0 0 240 157]{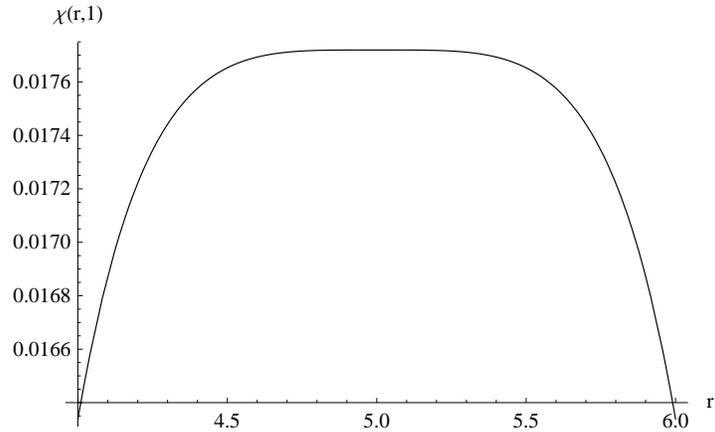}
  \caption{Eigenfunction for the nonlinear warp factor closed to the brane.}
   \label{nonlineareigenfunctiona1}
\end{figure}

The eigenfunction must satisfies
\begin{equation}
 \chi''(r)+\left(\frac{(r-5)}{(r-5)^{2}+9a^{2}}-\frac{5}{2}A'(r)\right)\chi'(r)+
\left(\frac{(r-5)^{2}+6a^{2}}{(r-5)^{2}+9a^{2}}\right)e^{A(r)}m^{2}\chi(r)=0.
\end{equation}
For $\left(\frac{(r-5)}{a}\right)\rightarrow 0$ we can expand the warp factor and its exponential around $r=5$. This
yields the equation
\begin{equation}
 \chi''(r)+\left(\frac{(r-5)}{9a^{2}}-\frac{5}{2}(r-5)\right)\chi'(r)+
\frac{2}{3}(r-5)^{2}m^{2}\chi(r)=0.
\end{equation}
The solution for this equation is again a product between a exponential function and the hypergeometric confluent function. Its plot is shown in the figure
(\ref{nonlineareigenfunctiona1}).

\begin{figure}
 \centering
 \includegraphics[scale=1.1,bb=0 0 240 157]{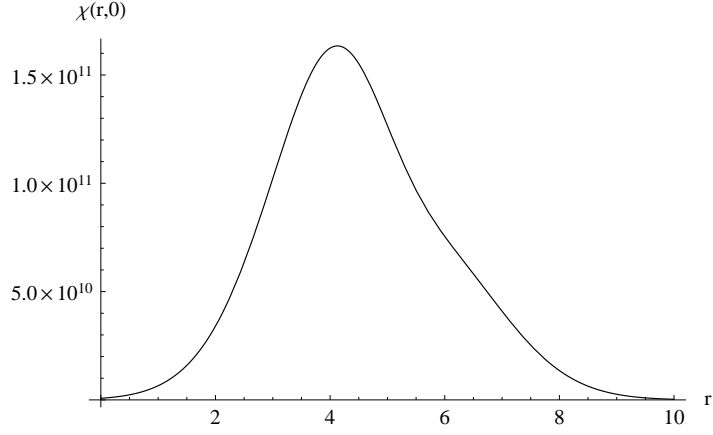}
  \caption{Eigenfunction for the nonlinear warp factor far from the the brane and or for $a=0$.}
 \label{nonlineareigenfunctiona0}
\end{figure}

For $\left(\frac{a}{r-5}\right)\rightarrow 0$ the eigenfunction satisfies
\begin{equation}
 \chi''(r)+\left(\frac{5(r-5)}{2}\right)\chi'(r)+
(r-5)^{2}\chi(r)=0
\end{equation}
whose solution is
\begin{equation}
 \chi(r)=e^{10r-r^{2}}\left(H\left(-\frac{4}{3},\frac{\sqrt{3}}{2}(r-5)\right)+M\left(\frac{2}{3},\frac{1}{2},\frac{\sqrt{3}}{2}(r-5)\right)\right)
\end{equation}
We have plotted the solution above in the figure (\ref{nonlineareigenfunctiona0}). Note that the function is well
defined on the brane but asymmetric in relation to the brane. Therefore, the eigenfunction for the singular conifold $(a=0)$ is localized on the brane. Besides,
since the eigenfunction vanishes at infinity, the eigenfunction for $a\neq 0$ is also localized.

\subsection{Massless modes}
\label{Massless mode}

Now let us turn our attention to the massless modes ( or Kaluza-Klein modes). Considering $m=0$ and $l=0$ (which is called \textit{s-wave}) \cite{Oda:2000zc}, the
radial equation (\ref{radialequation}) becomes
\begin{equation}
\label{masslessequation}
 \chi_{k}''(r)-\frac{1}{2}\left(4A'+B'+\frac{\alpha'}{\alpha}-\frac{\beta'}{\beta}\right)\chi'_{k}(r)=0.
\end{equation}
The constant function $\chi(r)=\chi_{0}$ satisfies the equation above. Thus, this solution is said to be localized
if its action is localized around the 3-brane, i.e., if its action have compact support. Since the equation
(\ref{masslessequation}) is a Sturm-Liouville equation and we are seeking localized functions that satisfy the
asymptotic condition
\begin{equation}
 \lim_{|r-5|\rightarrow \infty}\chi'(r)=0,
\end{equation}
we can find a spectra of eigenfunctions $(\phi(r))_{n}$ satisfying the orthonormal condition
\begin{equation}
 \int_{0}^{\infty}{e^{-\frac{3A(r)}{2}}\sqrt{\alpha\beta}*\phi_{n}(r)\phi_{m}(r)dr}=\delta_{nm}.
\end{equation}
Therefore, we can define the eigenfunction in flat space as
\begin{equation}
 \chi_{n}(r,a)=e^{-\frac{3A(r)}{4}}(\alpha(r,a)\beta(r,a))^{\frac{1}{4}}\phi_{n}(r).
\end{equation}
On the other hand, since
\begin{equation}
 \chi_{0}^{2}\int_{0}^{\infty}{e^{-\frac{3A(r)}{2}}\sqrt{\alpha\beta}dr}=1
\end{equation}
the zero-mode eigenfunction is
\begin{equation}
 \chi_{0}(r,a)=le^{-\frac{3A(r)}{2}}\sqrt{\alpha(r,a)\beta(r,a)},
\end{equation}
where
\begin{equation}
 l=\frac{1}{\int_{0}^{\infty}{e^{-\frac{3A(r)}{2}}\sqrt{\alpha\beta}}dr}.
\end{equation}

\begin{figure}
\centering
 \includegraphics[scale=1.1]{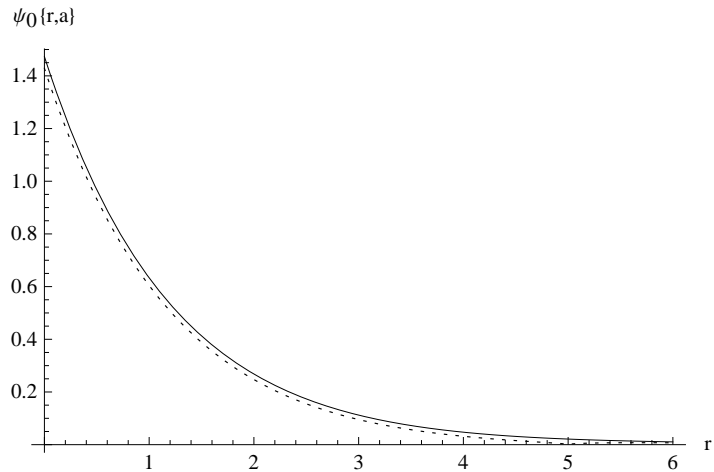}
\caption{Plot of the zero-mode eigenfunction for the linear warp factor. For any $a$ it has an exponential
decreasing behavior and thus it is localized (but not symmetric) around the brane.}
\label{masslesslinear}
\end{figure}

\begin{figure}
\centering
 \includegraphics[scale=1.1]{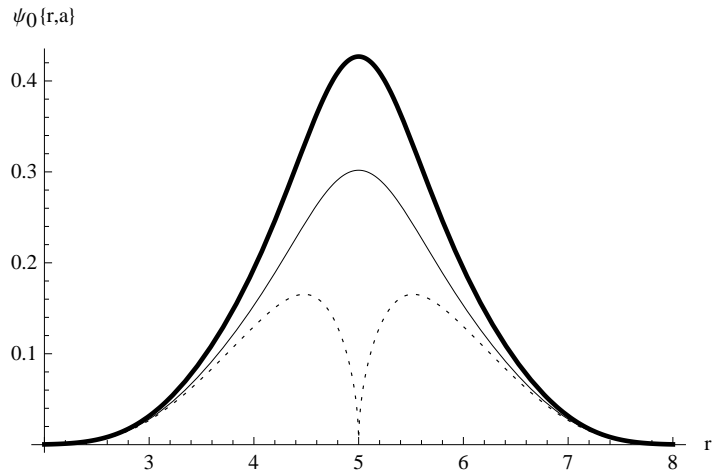}
\caption{Plot of the zero-mode eigenfunction for the nonlinear warp factor. The function has compact support even
for the singular conifold $a=0$ (dotted line). Note that this solution is localized and symmetric around the brane.}
\label{masslessnonlinear}
\end{figure}

Again, the eigenfunction is quite similar to that found in refs. \cite{Gherghetta:2000qi,Oda:2000zc} regardless the factor $(\alpha(r,a)\beta(r,a))^{\frac{1}{2}}$.
We have plotted the zero-mode $\psi_{0}(r,a)$ for some values of $a$ using the linear warp factor in the
fig. (\ref{masslesslinear}) and for the nonlinear warp factor in the fig. (\ref{masslessnonlinear}).
Since for every $a$ the function is integrable we can say that the massless field is localized on the brane even for the
singular conifold case.

\section{Conclusion and perspectives}
\label{Conclusions}

We have studied the localization of a scalar field on a 3-brane in a resolved conifold background
of six dimensions, builded from
a warped product between a 2-cycle of a resolved conifold and a 3-brane placed in the tip of the cone. We have
chosen a geometry such that when
the resolution parameter goes to zero or when we put the 3-brane far from the origin, we re-obtain the
well-studied string-like geometries.

The use of the resolved conifold as a transverse space brought a very nice feature: an extension to the
solution of the hierarchy problem using the resolution parameter $a$. Indeed, in the Randall-Sundrum and string-like
models, the ratio between the six-dimensional Planck mass $M_{6}$ and the four dimensional Planck mass $M_{4}$
depends only on the cosmological constant $\Lambda$ and on the tension of the brane $\mu$. Since this parameter has a purely geometric
origin it has opened the way to use another parameter-dependent manifolds like Eguchi-Hanson spaces, Taub-Nut, etc,
as transverse spaces. Furthermore, using parameter dependent transverse spaces we could study the evolution
and stability of the hierarchy between $M_{6}$ and $M_{4}$ through some mechanisms, for example, the conical transitions or the
geometrical flux.

For a linear warp factor, the bulk geometry asymptotically approaches to the $AdS_{6}$ space ($\Lambda <0$) which is
similar to the string-like geometry.
Further, the massive modes have a family of potentials parameterized by the resolution parameter.
For $a\geq 1$, the potential decreases exponentially and then it is not possible to have stable trapped states in
the brane.
As $a\rightarrow 0$, two asymmetric minima appear around the 3-brane put in the point $r=5$, but the potential
reaches its maximum value on the brane. The respective eigenfunction is localized around the deepest minimum
close to the brane. In the limit $a=0$, the potential well turns out
to be a infinite potential well in the brane and thus the eigenfunction is localized in the brane. Indeed, both for $\frac{a}{(r-5)}\rightarrow 0 $,
and for $\frac{(r-5)}{a}\rightarrow 0$ we obtain the well-known solution for string-like geometry \cite{Gherghetta:2000qi}.
Besides, the massless modes was localized even for the singular case. Indeed, the action for a constant solution
in this geometry has compact support and therefore is normalizable. The main difference between this conifold
geometry and the standard string-like one lies on the parameterization of the potentials by a geometric factor.
This parameterization provides the asymmetric minima potentials far from the brane. Therefore, the resolution parameter
could be used as a filter of fields that can be or not be localized on the brane.

Using the nonlinear warp factor $A(r)=\cosh(r)+\tanh(r)^{2}$ the scalar curvature diverges asymptotically
and so the
energy-momentum tensor is not restricted to points closed to brane which is in contrast to the string-like geometry.
For the scalar field, the massive modes are trapped on the 3-brane even for
the singular conifold.
For $a\geq 1$ the potential behaves like the standard volcano potential with a minimum on the brane. As
$a\rightarrow 0$ two more local
minima appear inside the global minimum and the minimum on the brane turns to be a maximum. Therefore, in the range
$0<a\leq 1$ is not possible localize the scalar field on the brane. In the same way that
for linear warp factor, for $a=0$ the potential
on the brane returns to be a potential well but now infinite. At last, the massless modes are localized for all values of
$a$. A difference between this geometry and the string-like one lies on the radial field solutions. Indeed,
for string-like geometries the usual field solutions are Bessel functions and in this work we have found confluent
hypergeometric functions that depend on two parameters and which are very sensible to variations on these parameters.
This was expected since the field equation depends on the mass $m$ and angular number $l$. Also, the field equation, through
the geometry, depends on the resolution parameter $a$.
We argue that using other parameter-dependent geometries we could find more general solutions and study its stability.

We also argue that the successful localization of massless modes in the singular conifold for both linear and nonlinear warp factor
might be related with the fact that in the conifold transitions, where we make singularities in the manifold, some fields become massless. Thus,
this closed relationship between conical singularity and the spectrum of massless fields provides the localization
of the massless modes in the singular geometry.

This work suggests many perspectives. Using the same geometry studied so far, we could study the localization of
other fields in this scenario, like the vector, gravitational and spinor fields. Since these fields have more degrees
of freedom than the scalar field the effects of asymmetry of the potential could be more relevant, for example, for
the resonance of fermionic modes. Another way it could be use another parameter-depend geometries like the deformed conifold or even
orbifold instead of the resolved conifold. Since these smooth conifolds are related with supergravity solutions
that near horizon behaves like $AdS_{5}\times S^{5}$ we could study the localization of fields where the transverse
space are well-known solutions like the Eguchi-Hanson spaces or the Klebanov-Strassler throat.
Furthermore, since we have parameterized the geometry we could study the stability of this geometry using some analytical
method like Ricci equation, where the variable of the flux would be the resolution parameter. Since there is a relation between
the resolved conifold geometry and the cigar-like geometry we could study the flux through a parameterized cigar geometry.
On the other hand, we could use the geometry bulk used here to study other problems as the small value of the cosmological constant in the
3-brane and the supersymmetry breakdown.

It worthwhile to mention that even though we have studied the behavior of the field in that geometry we have not said
how this geometry was generated. Hence, a next step could be deduce this geometry from some field and so to give a
physical meaning to the resolution parameter.

The authors would like to thank the Brazilian agencies FUNCAP, CNPq and CAPES for financial support.

\end{document}